%% file: main.tex
\documentclass[12pt]{article}

\usepackage[utf8]{inputenc}
\usepackage{hyperref} 
\usepackage{xcolor}
\usepackage{graphicx}
\usepackage{amsmath}
\usepackage{mathtools}
\usepackage{amssymb}
\usepackage{amsfonts}

\newcommand{\pd}{\partial}
\newcommand{\tr}{\mathop{\mathrm{tr}}}

\begin{document}

\begin{flushright}\footnotesize
\texttt{NORDITA-2020-051} \\
\vspace{0.6cm}
\end{flushright}

\begin{center}
{\Large\textbf{\mathversion{bold} Exact D7-brane embedding in\\ the Pilch-Warner background}
\par}

\vspace{0.8cm}

\textrm{Xinyi Chen-Lin\footnote{xinyitsenlin@gmail.com}, Amit Dekel\footnote{amit.dekel@gmail.com}}
\vspace{4mm}

\textit{Nordita\\
Roslagstullsbacken 23, SE-106 91 Stockholm, Sweden}

\vspace{5mm}

\textbf{Abstract} 
\vspace{5mm}

\begin{minipage}{13cm}
\input{abstract}
\end{minipage}

\end{center}

\newpage

\tableofcontents

\input{sections/introduction}

\input{sections/Dbranes}

\input{sections/D7}

\input{sections/action}

\input{sections/conclusions}

\input{acknowledgement}

\input{sections/appendix}

\bibliographystyle{nb}
\bibliography{refs}
\end{document}

%% file: abstract.tex
A new supersymmetric D7–brane embedding in the Pilch–Warner gravitational background is found exactly, by solving the supersymmetric condition. 
In the dual holographic picture, our setting corresponds to adding a quenched fundamental matter sector to  $\mathcal{N}=2^*$  super Yang–Mills theory, at zero temperature. We show that previous results in the same setting are missing the Wess-Zumino term in the D-brane action, and how our results complete the picture.

%% file: sections/introduction.tex
\section{Introduction}

The holographic principle promises to be a useful framework to tackle strongly coupled gauge theories by means of weakly coupled string theories. Its best known instance is the AdS/CFT duality, conceived by Maldacena in his seminar work \cite{Maldacena:1997re}. In particular, the duality involves a strongly coupled $SU(N)$ $\mathcal{N}=4$ SYM, which is a CFT, and the supergravity in the $AdS_5 \times S^5$ background. This is a very well understood case by now, and despite not describing any real world systems, AdS/CFT is being used as a theoretical laboratory to explore dynamics of real strongly coupled systems.

The most famous example is undoubtedly Quantum Chromodynamics (QCD). To compare $\mathcal{N}=4$ SYM with QCD, we must first add flavors to AdS/CFT. This means additional hypermultiplet sectors with fields in the fundamental representation of the gauge group $SU(N)$. The equivalent in the holographic picture is the insertion of probe D7-brane embeddings in $AdS_5 \times S^5$, see \cite{Karch:2002sh}. The fundamental matter fields, or quarks, arise from strings stretched between the D7 and the $N$ coincident D3-branes that generate the supergravity background. When the D7-branes and the D3-branes are separated, the quarks become massive. Significant work has been done in this direction, and we refer readers to the following reviews \cite{CasalderreySolana:2011us, Erdmenger:2007cm}.

In this paper, we study the flavor dynamics in a less symmetric theory called $\mathcal{N}=2^*$ SYM, via the holographic correspondence. This theory results from breaking the conformal invariance of $\mathcal{N}=4$ SYM by adding mass to its adjoint hypermultiplet. Consequently, the supersymmetries are also halved. Its holographic dual is known too, namely the Pilch-Warner supergravity \cite{Pilch:2000ue, Pilch:2003jg}. This geometry consists of a product space of a warped $AdS_5$ and a squashed $S^5$, and is asymptotically $AdS_5 \times S^5$ near its boundary. This instance of the holographic duality is a non-conformal extension of AdS/CFT, and has been extensively studied and tested in the last few years \cite{Buchel:2013id, Chen-Lin:2015xlh, Chen-Lin:2017pay, Russo:2019lgq}. It is therefore a natural step to extend the flavor sector in $\mathcal{N}=2^*$ theory. 

This problem was studied in \cite{Albash:2011nw}, where a perturbative solution to the D7-brane equation of motion was found, and an unexpected logarithmic divergence in the embedding profile was encountered. In their analysis, the authors argued that the Wess-Zumino Lagragian vanishes. In contrast, we show that such divergence does not arise if the D-brane couples to the Ramond-Ramond (RR) fluxes through Wess-Zumino. We provide an exact closed-form solution, obtained
by solving the supersymmetric condition imposed on the probe D7-brane embedding.

The present paper starts, in section \ref{sec:Dbranes}, by reviewing D-branes and outlining the strategy we use to find supersymmetric embeddings. Then, in section \ref{sec:D7brane}, we deal with our D7-brane in detail, and find the right configuration. We check that our solution satisfies the equation of motion and study the renormalized action in \ref{sec:action}. Finally we conclude by describing the implications of our results. The Pilch-Warner background is summarized in the appendix, including explicit forms of the RR potentials that we computed for completeness.

%% file: sections/Dbranes.tex
\section{D-branes}\label{sec:Dbranes}

\subsection{Kappa symmetry projector}
\input{sections/kappa_symmetry}

\subsection{Supersymmetric condition}
The condition for the D-brane configuration to be supersymmetric is that the kappa symmetry projector $\Gamma$ applied to the background Killing spinor $\epsilon$ fulfills
\footnote{The sign in front of the spinor is positive if mostly-plus metric is used, as for example in \cite{Skenderis:2002vf}.}:
\begin{equation} \label{eq:susyCondition}
 \Gamma \epsilon = - \epsilon.
\end{equation}

If we are to impose the supersymmetric condition on an ansatz, it will lead us to first order differential equations for the ansatz, which are easier to solve than the standard second order equations of motion from the D-brane action. Here we will outline our strategy to solve the supersymmetric condition under certain conditions.

Let us consider the Killing spinor with the following structure:
\begin{equation} \label{eq:epsilonDecomposition}
\epsilon = \mathcal{O} \mathcal{P} \epsilon_0,
\end{equation}
where $\mathcal{O}$ is an invertible operator and $\mathcal{P}$ is a projector satisying
\begin{equation}
    \mathcal{P} \epsilon_0 =  \epsilon_0,
\end{equation}
so that there exists a complementary projector satisying
\begin{equation}
    \bar{\mathcal{P}} \epsilon_0 =  0.
\end{equation}

Then, the supersymmetric condition
\begin{equation}
 \Gamma \mathcal{O} \mathcal{P} \epsilon_0 = - \mathcal{O} \mathcal{P} \epsilon_0
\end{equation}
implies the condition
\begin{align}\label{eq:susyCondition0}
&\bar{\mathcal{P}} \mathcal{O}^{-1} \Gamma \mathcal{O} \mathcal{P}  = 0.
\end{align}

If we find $n$ further projectors on the Killing spinor as necessary conditions for the supersymmetric condition to be fulfilled, then, it means that the D-brane configuration breaks $1/2^n$ copies of the background supersymmetry.

\subsection{Action}
The supersymmetric configuration to be found using the above-mentioned method must also be a solution of the D-brane equation of motion. For this purpose, let us state the world-volume action of a single D$p$-brane. It consists of the Dirac-Born-Infeld (DBI) and the Wess-Zumino (WZ) or Chern Simons terms\footnote{An anti-brane corresponds to a sign change in front of the Wess-Zumino term, \cite{Kruczenski:2003be}.},
\cite{Ammon:2015wua}:
\begin{equation} \label{eq:DbraneAction}
 S = S_{DBI} + S_{WZ},
\end{equation}
which are explicitly
\begin{align}
 S_{DBI} & = 
 -T_p \int_\mathcal{M} d^{p+1}\xi \, e^{-P[\Phi] } \sqrt{-\det (g+\mathcal{F})},\\
 S_{WZ} & =
 T_p\int _\mathcal{M} \sum_n e^{\mathcal{F}}\wedge P[C_{(n+1)}],
\end{align}
where $\xi$ are the coordinates for the worldvolume manifold $\mathcal{M}$, $g$ is the worldvolume metric (in string frame), $P[\cdot]$ denotes the pullback from the target space, $\Phi$ is the dilaton, $C_{(n)}$ are the RR forms, and $\mathcal{F}$ is defined in \eqref{eq:defF}. Finally, the couplings, in terms of the string length $l_s$ and the string coupling constant $g_s$, are:
\begin{equation}
 T_{s} = \dfrac{1}{2\pi l_s^2}, \quad T_p = \dfrac{1}{g_s} T_s (2\pi l_s)^{1-p}.
\end{equation}

%% file: sections/kappa_symmetry.tex
For any D-brane configuration there is an associated kappa symmetry projector, which, in Minkowski signature, is given by\cite{Skenderis:2002vf}:
\begin{align}
d^{p+1} \xi \, \Gamma = - \dfrac{ e^{\mathcal{F}}\wedge X|_{\text{Vol}}}{\sqrt{-\det  (g+\mathcal{F})}},
\end{align}
where
\begin{align}
X = \bigoplus_n \gamma_{(2n)} \mathcal{K}^n \mathcal{I},
\end{align}
and $|_{\text{Vol}}$ indicates projection to the volume form.
The operators $\mathcal{K}$ and $\mathcal{I}$ act on a spinor $\psi$:
\begin{equation}
 \mathcal{K} \psi = \psi^* , \quad \mathcal{I} \psi = -i \psi.
\end{equation}
We also have
\begin{align}
\gamma_{(n)} = \frac{1}{n !}d\xi^{i_n}\wedge ... \wedge d\xi^{i_1} P[\gamma_{i_1...i_n}],
\end{align}
built from the pullback $P[\cdot]$ of the gamma matrices in the curved target space; and 
\begin{equation}\label{eq:defF}
 \mathcal{F} \equiv \frac{1}{T_s} F + P[B_{(2)}], 
\end{equation}
with $F$ being the worldvolume field strength, and $B_{(2)}$ the NSNS 2-form.

The kappa symmetry projector satisfies the traceless and idempotent conditions.

%% file: sections/D7.tex
\section{D7-brane}\label{sec:D7brane}

The holographic dictionary for $N_f$ flavors of quarks in a four-dimensional $SU(N)$ SYM theory is a set of $N_f$ D7-branes in the ten-dimensional supergravity dual. We work in the probe limit, when $N_f/N \rightarrow 0$, meaning the additional branes do not back-react on the background geometry. Furthermore, at this limit, the Landau pole that can potentially develop in the dual theory is strongly suppressed, \cite{CasalderreySolana:2011us}. We study the simplest setting, with $N_f=1$ probe brane \footnote{For many coincident branes in the probe limit, they are non-interacting, hence the non-abelian action that describes them reduces to $N_f$ copies of the abelian action.}.

Our D7-brane embedding wraps the warped $AdS_5$ and the three-dimensional ellipsoid of the deformed $S^5$ of the Pilch-Warner metric. Furthermore, our D7-brane carries no charge, hence no worldvolume gauge field: $F = 0$. This is the equivalent setting studied in \cite{Karch:2002sh} for $AdS_5 \times S^5$, which our configuration will reduce to, near the boundary.

Let us consider the D7-brane worldvolume, induced from the target space with 
\begin{equation}\label{eq:ansatz}
 \theta = \theta(c), \quad \phi=\phi_0\equiv\frac{(2 n + 1)\pi}{2}.
\end{equation}
Our particular choice of $\phi_0$ simplifies our problem, because
\begin{equation}
 P[B_{(2)}] = 0.
\end{equation}
The induced metric from \eqref{eq:PWmetric} with $d\theta = \theta'(c) dc$ and $d\phi=0$ is thus:
\begin{align}\label{eq:D7metric}
ds_{D7}^2 =
v_x^2 dx_\mu dx^\mu 
- (v_c^2 +v_\theta^2 \theta'(c)^2)\, dc^2 - v_1^2 \sigma_1^2 - v_2^2 (\sigma_2^2 + \sigma_3^2),
\end{align}
where we used lower case $v$ to denote the pullback of the target space vielbeins.

%%%%%%%%%%%%%%%%%%%%%%%%%%%%%%%%%%%%%%%%%%%%%%%%%%%%%%%%%%%%%%%%%%%%%%%%%%%%%%%%%%%%%%%%%%%%%%%%%
\subsection{Kappa symmetry projector}

The kappa symmetry projector for our configuration is:
\begin{align}
\Gamma = - \dfrac{ \gamma_{(8)} \mathcal{I} }{\sqrt{-\det g}},
\end{align}
with
\begin{align}
 \gamma_{(8)} = - v_x^4 v_1 v_2^2 \Gamma_{1 2 3 4 7 8 9}( v_c \Gamma_5 +  v_{\theta} \theta'(c) \Gamma_6), 
\end{align}
where we used capital gammas to denote the gamma matrices in the local frame; see appendix \ref{sec:localframe}.

The projector can be further simplified by combining it with the chirality condition, which for the mostly-minus metric convention is: 
\begin{equation}
 \Gamma_{11} \epsilon = -\epsilon, \quad 
 \Gamma_{11} \equiv \Gamma_{12345678910}.
\end{equation}
Then, the supersymmetric condition \eqref{eq:susyCondition} becomes:
\begin{equation}
 \Gamma_{11} \Gamma \epsilon = \epsilon,
\end{equation}
and 
\begin{align} \label{eq:newProjector}
  \mathcal{P}' \equiv \Gamma_{11} \Gamma  = \dfrac{1}{\sqrt{1+\xi^2}}(1- i \xi  \Gamma_{510}), \quad 
   \xi \equiv  \dfrac{v_\theta}{v_c} \theta'(c),
\end{align}
where we have applied $\mathcal{I} \epsilon = -i\epsilon$ and $\Gamma_{610} \epsilon = -i \epsilon$. The latter identity is due to $\mathcal{P}_- \epsilon = 0$, which is straightforward to show, and the projector is defined in \eqref{eq:projectors}.

%%%%%%%%%%%%%%%%%%%%%%%%%%%%%%%%%%%%%%%%%%%%%%%%%%%%%%%%%%%%%%%%%%%%%%%%%%%%%%%%%%%%%%%%%%%%%%%%%
\subsection{Supersymmetric condition}

The Killing spinor \eqref{eq:KillingSpinor} can be decomposed in the form \eqref{eq:epsilonDecomposition}, where the invertible operator and its inverse, and the projector are:
\begin{align}
 \mathcal{O} &= \exp{\left(\frac{\alpha}{2}\Gamma_{56} \right)} \exp{\left(-\frac{\phi}{2}\, \Gamma_{610} \right)} \exp{\left(\frac{\beta}{2}\Gamma_{710} \mathcal{K} \right)}, \\
 \mathcal{O}^{-1} &=  \exp{\left(-\frac{\beta}{2}\Gamma_{710} \mathcal{K} \right)} 
 \exp{\left(\frac{\phi}{2}\, \Gamma_{610} \right)} 
 \exp{\left(-\frac{\alpha}{2}\Gamma_{56} \right)},\\
 \mathcal{P} &= \Pi_+ \mathcal{P}_+.
\end{align}

The kappa symmetry projector contains the operator $\mathcal{I}$, which can be replaced as follows (see notation in \ref{sec:KillingSpinor}):
\begin{equation}
 \mathcal{I}\eta =-i \eta = \Gamma_{610} \eta,
\end{equation}
where we used $\mathcal{P}_- \eta =0$ in the last step. 

Furthermore, since the projectors in \eqref{eq:projectors} commute, it is sufficient to study the supersymmetric condition \eqref{eq:susyCondition0} with one set of them, for example, with $\Pi_\pm$:
\begin{equation}
 \Pi_{-} \mathcal{O}^{-1} \gamma_{(8)} \mathcal{O} \Gamma_{610} \Pi_{+}  = 0.
\end{equation} 
After manipulating the gamma matrices, it gives:
\begin{equation}
i v_1 v_2^2 v_x^4 \Pi_- \Gamma_{6 8 9 10} \mathcal{K} \sin\beta \left(v_c \sin\alpha - v_\theta \cos\alpha \, \theta'(c)\right) = 0.
\end{equation}
Therefore, the condition our configuration must satisfy in order to preserve supersymmetry is:
\begin{equation}\label{eq:susyConditionTheta}
 \theta'(c) = \dfrac{v_c}{v_\theta} \tan\alpha = \dfrac{c \, \tan\theta(c)}{c^2-1} .
\end{equation}

The projector \eqref{eq:newProjector} at the solution \eqref{eq:susyConditionTheta} is simply
\begin{equation}
\mathcal{P}' = \cos \alpha - i \sin \alpha \, \Gamma_{510},
\end{equation}
since $ \xi =  \tan \alpha $.
No more projectors are found, therefore, ours is a 1/2-BPS embedding.

\input{sections/solution.tex}

%% file: sections/solution.tex
\subsection{Solution}\label{sec:solution}

The solution to the differential equation \eqref{eq:susyConditionTheta} is:
\begin{equation}\label{eq:susyConditionSolution}
\boxed{\sin\theta(c) = L \sqrt{c^2-1}; \quad 1 < c \leq \sqrt{1+L^{-2}}},
\end{equation}
where $L$ is an integration constant. As we will show below, it is the asymptotic separation of the D7-brane from the stack of the D3-branes, in the units of the spherical radius\footnote{The spherical part of our metric \eqref{eq:PWmetric} is multiplied by $R^2$.} $R$, namely $L$ above is really $ L/R$.
We have set $R=1$ so far. The upper bound of $c$ is set by the maximum of the sine.

Near the boundary, $c \approx 1 + z^2/2$, the solution behaves as
\begin{equation} \label{eq:thetaExpanded}
 \theta(z) \approx L \, z + \left(\frac{L}{8} +\frac{L^3}{6} \right) \, z^3 + O(z^5).
\end{equation} 
Moreover, keeping only the leading order of the large $L$ expansion, our solution reduces to the one found in the $AdS_5 \times S^5$ background, see \cite{Karch:2002sh} and \cite{Karch:2005ms}, i.e.
\begin{equation}
 \sin\theta(z)_\text{AdS} = L z,
\end{equation}
with the asymptotic expansion
\begin{equation}
\theta(z)_\text{AdS} \approx L z + \frac{L^3}{6} z^3 + O(z^5).
\end{equation}
The $L\gg 1$ limit ensures the upper bound for $c$ in \eqref{eq:susyConditionSolution} to reduce to the one from the AdS solution, namely $z_\text{max}=1/L$.

As \cite{Karch:2005ms} explains, in the flat embedding space limit, this embedding describes a planar D-brane located at a constant distance $L$ away from the stack of $N$ D3-branes:
\begin{equation}
 L = \lim_{z \rightarrow 0 } \frac{R}{z} \sin\theta(z) = R \, L/R,
\end{equation}
where we explicitly stated $R$. Furthermore, this separation is proportional to the quark mass $m$:
\begin{equation}
 L = 2 \pi l_s^2 m.
\end{equation}

\begin{figure}[t!]
\begin{center}
\includegraphics[width=0.6\textwidth]{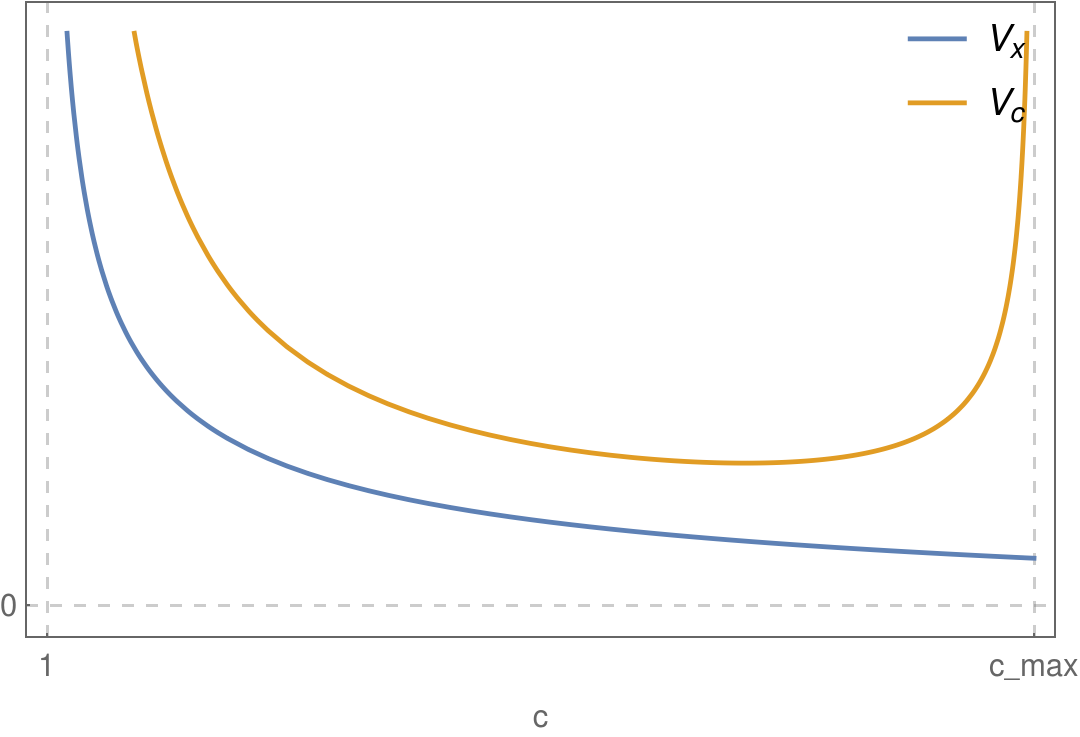}
\end{center}
\vspace{0.05mm}
\begin{center}
\includegraphics[width=0.6\textwidth]{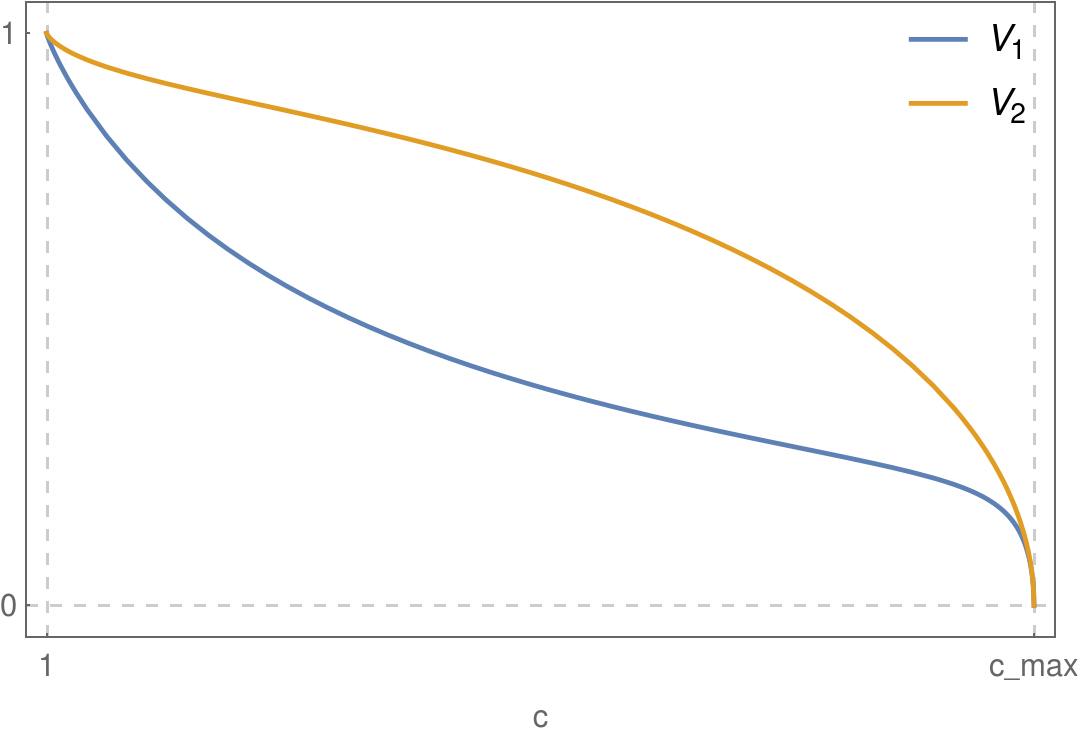}
\end{center}
\caption{\label{fig:vielbeins} The vielbeins of the induced metric for the allowed values of $c$.}
\end{figure}

Figures in \ref{fig:vielbeins} show the vielbeins of the induced metric at the solution, from which we learn how the geometry of the embedding looks like at different values of $c$. First, observe the divergence at the horizon $c_\text{max}=\sqrt{1+L^{-2}}$. This is the location of the well-known enhançon locus, at $\theta = \pi/2$, see \cite{Buchel:2000cn} and \cite{Evans:2000ct}. The spheroid is undeformed at the boundary $c=1$, and becomes squashed until it vanishes at the enhançon. 

%% file: sections/action.tex
\section{Action}\label{sec:action}

The D7-brane action \eqref{eq:DbraneAction} for our configuration \eqref{eq:ansatz} can be written as:
\begin{align} \label{eq:ActionWithTheta'}
 S  = & -T_7 \int_\mathcal{M} d^8\xi \, e^{-P[\Phi] } v_x^4 v_c v_1 v_2^2 \sqrt{1+\frac{v_\theta^2}{v_c^2}\theta'(c)^2}  \nonumber \\
      & + T_7\int _\mathcal{M} P[C_{(8)}],
\end{align}
or more explicitly, using the solution \eqref{eq:susyConditionTheta}, as:
\begin{align}\label{eq:ActionWithTheta}
 S = & -T_7 \int_\mathcal{M} d^8\xi \, \dfrac{c A(c) \cos^3\theta (c) \sqrt{X_1(c, \theta(c))}}{\left(c^2-1\right)^3} \sqrt{1+ c A(c) \tan^2\theta(c)} \nonumber \\
     & +T_7\int _\mathcal{M} d^8\xi \, \dfrac{A(c)^2 \cos^4\theta(c)}{4 \left(c^2-1\right)^2}.
\end{align}
The explicit form of the Wess-Zumino term is calculated below.

\subsection{Wess-Zumino term}
The $P[C_{(8)}]$ term was deemed vanishing in \cite{Albash:2011nw} and \cite{Evans:2005ti}. Their argument did not consider the dilaton factor in the string frame that affects the Hodge star operation while deriving $C_{(8)}$, which in our scenario is
\begin{equation}
 dC_{(8)} = \ast dC_{(0)}.
\end{equation}
The dilaton term from the string frame effectively cancels the factor that vanishes at $\phi_0$, leading to a finite value for the pullback of this potential. We decided to compute it explicitly, and the full result is given in the appendix \ref{sec:backgroundFields}. We can quickly see that it is non-zero for our ansatz for $\phi$ in \eqref{eq:ansatz}. However, $P[C_{(8)}]$ term can be much simpler, as we will see now. First, we can show that:
\begin{equation}\label{eq:C8id}
 [d C_{(8)}]_{\phi_0} = d [C_{(8)}]_{\phi_0}.
\end{equation}
The left-hand-side is simply
\begin{equation}
 [d C_{(8)}]_{\phi_0}  = \dfrac{A^2 \sin\theta \cos^3(\theta)}{\left(c^2-1\right)^2} 
\sigma_1 \wedge \sigma_2 \wedge \sigma_3 \wedge dc  \wedge dx_0 \wedge dx_1 \wedge dx_2 \wedge dx_3 \wedge d\theta,
\end{equation}
and, via \eqref{eq:C8id}, we can integrate the above expression over $\theta$ and obtain:
\begin{equation}
[C_{(8)}]_{\phi_0} = \dfrac{A^2 \cos^4\theta}{4 \left(c^2-1\right)^2} \sigma_1 \wedge \sigma_2 \wedge \sigma_3 \wedge dc \wedge dx_0 \wedge dx_1 \wedge dx_2 \wedge dx_3.
\end{equation}
The full pullback is obtained by just replacing $\theta$ by $\theta(c)$.

%%%%%%%%%%%%%%%%%%%%%%%%%%%%%%%%%%%%%%%%%%%%%%%%%%%%%%%%%%%%%%%%%%%%%%%%%%%%%%%%%%%%%%%%%%%%%%%%%

\subsection{Equation of motion}

As a consistency check for our results, the equation of motion from the action \eqref{eq:ActionWithTheta'} is fulfilled with the solution \eqref{eq:susyConditionTheta}. In particular, 
\begin{align}\label{eq:eom}
-\left.EL[\mathcal{L}_{DBI}]\right|_\text{solution} = EL[\mathcal{L}_{WZ}] = \dfrac{A^2 \sin\theta \cos^3(\theta)}{\left(c^2-1\right)^2},
\end{align}
where $EL[\cdot]$ is the Euler-Lagrange operator:
\begin{equation}
 EL[\mathcal{L}] = 
 \left(\dfrac{\pd }{\pd \theta(c)} -\dfrac{\pd }{\pd c}  \dfrac{\pd }{\pd \theta'(c)} \right) \mathcal{L}.
\end{equation}
Therefore, \eqref{eq:eom} is another proof for the non-vanishing WZ term.

\subsection{Holographic renormalization}
The fully explicit on-shell action evaluated at the solution \eqref{eq:susyConditionSolution} is:
\begin{align}\label{eq:ActionAtSolution}
 S_\text{reg} = -T_7 V \int_{1 + \epsilon^2/2}^{c_\text{max}} d c \, 
 &\left[-\frac{c A(c)}{\left(c^2-1\right)^3} -\frac{A(c)^2}{4 \left(c^2-1\right)^2} \right. \nonumber\\
 &-\frac{L^2 A(c) \left(\left(c^2+1\right) A(c)-4 c\right)}{2 \left(c^2-1\right)^2} \nonumber\\
 &+\left.\frac{L^4 A(c) \left(3 c^2 A(c)+A(c)-4 c\right)}{4 \left(c^2-1\right)}\right],
\end{align}
where $c_\text{max}$ is the upper bound shown in \eqref{eq:susyConditionSolution}, and $V$ denotes the volume of the 4-dimensional Minkowski space times the 3-sphere. Notice that the action is divergent near the boundary, thus, we regularized it by adding a regulator $\epsilon>0$ in the lower integration limit. Let us also introduce the following notation regarding the integration limits:
\begin{equation}
  S_\text{reg} \equiv S_\text{IR}-S_\text{UV}.
\end{equation}

From the exact integrations in the appendix \ref{sec:integrate-action}, we can identify the divergent terms:
\begin{align} \label{eq:Sdiv}
 S_\text{div} = T_7 V 
        \left[ \frac{1}{4 \epsilon ^4} +\frac{1+\log \left(\epsilon/2\right)}{2 \epsilon ^2}-\frac{L^2}{2 \epsilon ^2}-\frac{ \log ^2\left(\epsilon/2\right)}{4}+\frac{\log (\epsilon )}{8} \right],
\end{align}
which come from the $L^0$ and $L^2$ terms of \eqref{eq:ActionAtSolution}. 

In holographic renormalization, the counterterms must be covariant and local on the regulator hypersurface. They do not only subtract the divergences, but also finite terms from both IR and UV regions, such that the final action vanishes, as required by supersymmetry. For general asymptotic AdS spaces and for the D7-brane in particular, the counterterms are derived in \cite{Karch:2005ms} (we do not copy the ones with curvature below):
\begin{align} \label{eq:Ls}
& L_{1}=-\frac{1}{4} \sqrt{\gamma} \nonumber\\
& L_{4}=\frac{1}{2} \sqrt{\gamma} \theta(\epsilon)^2 \nonumber\\
& L_{f}= -\frac{5}{12}\sqrt{\gamma} \theta(\epsilon)^4,
\end{align}
where $\gamma$ here denotes the determinant of the regulator hypersurface metric. 
However, for our case, these counterterms do not apply. Our geometry is seemingly not covered by this general study, due to the logarithmic divergence appearing already at the next-to-leading order $\epsilon$-expansion for the metric, coming from $A(c)$. The explanation is that the 10-dimensional Pilch-Warner background is uplifted from the 5-dimensional supergravity solution, and indeed the logarithmic divergence mentioned comes from fields in the 5-dimensional theory, not from the 5-dimensional metric; see \cite{Bobev:2013cja}. If we were to find covariant counterterms, these would be written in terms of these lower dimensional fields too.

Let us be concerned about the counterterms that cancel exactly $S_\text{div}$, for now. We propose:
\begin{equation}\label{eq:counterterms}
 S_\text{CT, UV} =  T_7 V \left[ 
  \frac{1}{4} \sqrt{\gamma} f(\epsilon)
   -\frac{1}{2} \sqrt{\gamma} \theta (\epsilon)^2 + \frac{1}{6} \sqrt{\gamma} \theta (\epsilon)^4
   \right],
\end{equation}
where the scalar field expansion is found in \eqref{eq:thetaExpanded}, and we defined
\begin{align}
 \sqrt{\gamma} &= \frac{1}{\epsilon^4}-\frac{1}{2 \epsilon ^2},\\
 f(\epsilon) &= 1 + 12 \alpha(\epsilon) + 6 \,(1 + 8 \alpha(\epsilon )) \chi(\epsilon ) ^2+14 \chi(\epsilon )^4, \label{eq:f}
\end{align}
where $\alpha(\epsilon)$ and $\chi(\epsilon)$ are scalar fields living in the 5-dimensional supergravity; we follow the notation in \cite{Albash:2011nw}, and our $A$ is their $\rho^6$.
The expression $f(\epsilon)$ is obtained from the numerator of \eqref{eq:I0} with the replacements 
\begin{align}
c &= \cosh(2 \, \chi) \approx 1 + 2 \chi ^2 + \frac{2}{3}\chi ^4,\\
A &= \exp{(6 \, \alpha)}\approx 1 + 6 \alpha + 18 \alpha ^2. 
\end{align}
From $c = 1 + \epsilon^2/2$ and the asymptotic expansions of $A$ in \eqref{eq:expandA}, we know that $\chi \propto \epsilon$ and $\alpha \propto \epsilon^2$ at the leading order. Then, for $f(\epsilon)$ we keep only the terms up to order $\epsilon^4$.  

In \eqref{eq:counterterms}, there is also a finite counterterm in terms of the quartic power of the scalar field. This is there to cancel the finite term introduced by the counterterm quadratic in the scalar field. 

\begin{figure}[t]
\begin{center}
\includegraphics[width=0.6\textwidth]{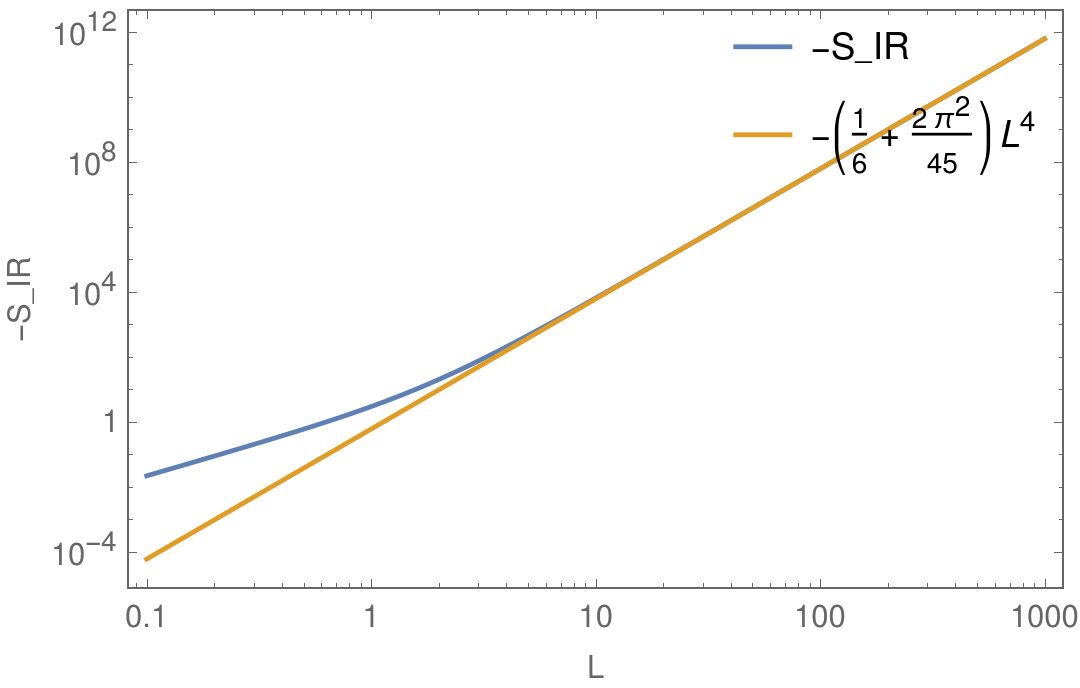}
\end{center}
\caption{\label{fig:SIR} This is a double logarithmic plot of the negative of the IR contribution to the action, $S_\text{IR}$, and its limiting behavior at large $L$.}
\end{figure}

The IR contribution to the action, $S_\text{IR}$, obtained by evaluating the integrals in \ref{sec:integrate-action} at $c=\sqrt{1+L^{-2}}$, are highly non-trivial functions of $L$, see figure \ref{fig:SIR}.
We are not able to find covariant counterterms to cancel them, and think it is not possible. 
As for the supergravity side, this is a finite term ambiguity in the renormalization scheme. 
However, from the field theory point of view, if we cannot write the counterterm in the covariant form, it would imply a non-vanishing action, and hence, a non-vanishing chiral condensate. 
It is known for $\mathcal{N}=1$ SYM theory that the chiral symmetry spontaneously breaks; see for example \cite{Bergner:2014saa}. Since our field theory also has one copy of supersymmetry, it is then possible that the condensate could be non-zero. We will leave a more detailed study of the dual field theory interpretation for future works.

As our last consistency check of our findings, let us study the large $L$ limit in the subsection below. 

\subsubsection{Large L}
As we discussed in \ref{sec:solution}, our embedding matches with the one in the $AdS_5 \times S^5$ background in the near-boundary expansion and at the large $L$ limit. 
The large $L$-expansion of the action is found in \ref{sec:IRlimit}. 
The finite terms from $S_\text{UV}$ are shown in \ref{sec:UVlimit}.
Finally, at the leading order in $L$, the total finite term is:
\begin{equation} 
S_\text{IR}-S_\text{UV}+S_\text{CT,UV} = - T_7 V \frac{1}{4} L^4 +O(L^2 \log L)
\end{equation}
Taking this term into account, the full counterterm action at the large $L$ limit is:
\begin{equation}
 S_\text{CT} =  T_7 V \left[ 
  \frac{1}{4} \sqrt{\gamma} f(\epsilon)
   -\frac{1}{2} \sqrt{\gamma} \theta (\epsilon)^2 + \frac{5}{12} \sqrt{\gamma} \theta (\epsilon)^4
   \right], \quad (L\gg 1).
\end{equation}
This indeed matches with the general counterterms \eqref{eq:Ls} if we set the fields in \eqref{eq:f} to zero.

The renormalized action, defined as
\begin{equation}
 S_\text{ren} = \lim_{\epsilon\rightarrow 0} (S_\text{reg}-S_\text{CT}),
\end{equation}
is hence exactly zero at the large $L$ limit. Consequently, the chiral condensate, sourced by $L$, also vanishes:
\begin{equation} 
\langle O \rangle = \frac{\delta S_\text{ren}}{\delta L} = 0. 
\end{equation}

%% file: sections/conclusions.tex
\section{Conclusion}

In this paper, we found an exact half-BPS D7-brane embedding in the Pilch-Warner background, by solving the supersymmetric condition. In the field theory side, this setting corresponds to adding a quark sector in the $\mathcal{N}=2^*$ SYM at zero temperature. Because the embedding breaks half of the background supersymmetries, the dual theory has a remaining $\mathcal{N}=1$ supersymmetry left.

We demonstrated that the pullback of $C_{(8)}$ for values of $\phi$ that are odd fractions of $\pi$ is non-zero. This is important since previous papers such as \cite{Albash:2011nw} and \cite{Evans:2005ti}\footnote{Their $\phi$ is shifted compared to ours, hence their sine is our cosine, and vice versa.} argued that the pullback of $C_{(8)}$ is vanishing while using the Einstein metric frame. In particular, the embedding in \cite{Albash:2011nw} differs from ours only in the WZ term, and their result has an extra logarithmic divergence. The D7-brane embedding of \cite{Evans:2005ti} is quite different from ours, as it wraps non-trivially the deformed sphere. The missing WZ term in principle contributes to their IR potential, hence potentially affecting their conclusion too. Our supersymmetric condition method does not require the D-brane Lagrangian, so that the fulfilment of the equation of motion at the solution provides a strong proof that the WZ term is there, besides our explicit computations of the fluxes. We can conclude also that the string frame is the right metric frame to use in the D-brane analysis, instead of the Einstein frame.

A rather surprising result we found is that the renormalized action for our embedding is a non-trivial function of the quark mass.
As a consistency check, our D7-brane configuration reduces to the known solution in the AdS/CFT case \cite{Karch:2005ms}, close to the boundary of the Pilch-Warner geometry. At this limit, the renormalized action indeed vanishes, and hence we can conclude that the chiral condensate is zero. It is definitely interesting to understand the general case, and see whether the dual $\mathcal{N}=1$ theory has indeed a non-vanishing condensate or not. It is suggested in the literature that chiral symmetry in zero temperature $\mathcal{N}=1$ SYM can be spontaneously broken. We will leave this investigation for future works.

As another possible future work, the fluctuations around our D-brane, corresponding to the meson spectrum on the field theory side, could be studied.
It would also be interesting to consider the scenario of a non-vanishing gauge field on the deformed sphere. For example, in $AdS_5 \times S^5$, \cite{Kruczenski:2003be} studied mesons that carry angular momentum. This scenario has also been studied by \cite{Karch:2015vra} for the global $AdS_5 \times S^5$, where different topological inequivalent solutions were found. One could also review the thermal case studied in \cite{Albash:2011dq}, for which the geometry was derived in \cite{Buchel:2003ah}.

%% file: acknowledgement.tex
\subsection*{Acknowledgements}
We would like to thank K.~Zarembo for reviewing the manuscript and useful comments.
This work was supported by the Marie Curie network GATIS of the European Union's FP7 Programme under REA Grant Agreement No 317089 and by the Knut and Alice Wallenberg Foundation.

%% file: sections/appendix.tex
\appendix

\input{sections/pilchwarner}

\input{sections/integrate_action}

%% file: sections/pilchwarner.tex
\section{Pilch-Warner supergravity}\label{sec:PWB}
The Pilch-Warner solution to the type II supergravity equations was originally found in \cite{Pilch:2000ue}. 
In this section, let us review its metric, the background fields and the Killing spinors, the latter first derived in \cite{Pilch:2003jg}. 

\subsection{Metric}\label{sec:metric}
\input{sections/metric}

\subsection{Background fields}\label{sec:backgroundFields}
\input{sections/background_fields}

\subsection{Killing spinor}\label{sec:KillingSpinor}
\input{sections/killing_spinor}

%% file: sections/metric.tex
We parametrize the ten-dimensional Pilch-Warner metric in the following way:
\begin{align}\label{eq:PWmetric}
ds^2 =
V_x^2 dx_\mu dx^\mu -\left(
V_c^2 dc^2 + V_\theta^2 d\theta^2 + V_1^2 \sigma_1^2 + V_2^2 (\sigma_2^2 + \sigma_3^2) + V_\phi^2 d\phi^2\right),
\end{align}
with $\mu=1,\ldots,4$ and the coordinates: $c\in(1, \infty), \, \theta \in [0, \pi/2], \, \phi \in [0, 2\pi]$. 

The various coefficients are functions of $c$, $\theta$ and $\phi$, where the dependence on the latter comes only from the dilaton $\Phi$ prefactor\footnote{The dilaton factor comes from the fact that we are using the metric in the string frame. In the Pilch-Warner literature, often the metric in the Einstein frame is shown. Both frames are related by a general conformal transformation, i.e. $ds^2_\text{Einstein} = e^{-\Phi / 2} ds^2_\text{string}$.}, with its explicit form shown in the next subsection. 
The coefficients are given by:
\begin{align}\label{eq:PWvielbeins}
V_x &=e^{-\Phi/4} \frac{c^{1/8} A^{1/4} X_1^{1/8} X_2^{1/8}}{(c^2 - 1)^{1/2}},\nonumber\\
V_c &= e^{-\Phi/4}\frac{c^{1/8}X_1^{1/8} X_2^{1/8}}{A^{3/4} (c^2-1)},\nonumber\\
V_\theta &= e^{-\Phi/4}\frac{X_1^{1/8} X_2^{1/8}}{c^{3/8}A^{1/4}},\nonumber\\
V_1 &= e^{-\Phi/4}\frac{A^{1/4}X_1^{1/8} }{c^{3/8}X_2^{3/8}} \cos\theta,\nonumber\\
V_2 &= e^{-\Phi/4}\frac{c^{1/8}A^{1/4}X_2^{1/8} }{X_1^{3/8}} \cos\theta, \nonumber\\
V_\phi &= e^{-\Phi/4}\frac{c^{1/8}X_1^{1/8} }{A^{1/4}X_2^{3/8}} \sin\theta,
\end{align}
and
\begin{align}
X_1 &=  \cos^2\theta + cA  \sin^2\theta,\nonumber\\
X_2 &= c \cos^2\theta + A  \sin^2\theta, \nonumber\\
A &= c+(c^2 -1)\frac{1}{2}\log\left(\frac{c-1}{c+1}\right).
\end{align}

The deformed 3-sphere is parametrized by the $SU(2)$ left invariant forms, i.e. the Maurer-Cartan forms:
\begin{equation}
\sigma_i = \tr(g^{-1}\tau_i d g), \quad i = 1,2,3
\end{equation}
where $\tau_i$ are the Pauli matrices and $g$ is a group element of $SU(2)$. The 1-forms satisfy the relation\footnote{Other conventions might introduce an extra global sign, for example in \cite{Buchel:2000cn}.}
\begin{equation}
 d\sigma_i  = \epsilon_{i j k} \sigma_j \wedge \sigma_k.
\end{equation}
We do not need to explicitly parametrize these forms for the purpose of the present paper; however, for the interested readers, there is an example using Euler angles in the appendix of \cite{Chen-Lin:2015xlh}.

\subsection{Local frame}\label{sec:localframe}
The non-coordinate basis, also known as the local frame, is specified by the Minkowski metric $\eta_{a b}$. It is related to the curved space metric $G_{M N}$ via vielbeins, according to:
\begin{equation*}
 G_{M N} = e^a_M e^b_N \eta_{a b}.
\end{equation*}
In our case, the metric \eqref{eq:PWmetric} is diagonal\footnote{Once we use an explicit parametrization for the deformed sphere, the vielbeins are not diagonal anymore.}, hence the vielbeins and the inverse vielbeins are precisely the coefficients \eqref{eq:PWvielbeins} and its inverse, respectively.

When we handle the curved-space gamma matrices $\gamma_{M}$, we will go to the local frame, in order to use the constant $\Gamma_a$ matrices:
\begin{equation*}
 \gamma_{M} = e^a_M \Gamma_{a}.
\end{equation*}

\subsection{The near-boundary geometry}

The boundary of the Pilch-Warner geometry is located at $c = 1$. 
Close to the boundary, $c \approx 1 + \frac{z^2}{2}$, with $z$ small, we recover the $AdS_5 \times S^5$ geometry in Poincare coordinates and the Hopf parametrization for $S^5$:
\begin{equation}\label{eq:AdS5xS5metric}
ds^2=\dfrac{dx^{\mu } dx_{\mu}-dz^2}{z^2}-\left(d\theta^2+\cos^2\theta \left(\sigma_1^2+\sigma_2^2+\sigma_3^2\right)+\sin^2\theta \,d\phi^2 \right).
\end{equation}

%% file: sections/background_fields.tex
The Pilch-Warner solution has non-trivial background fields. Following the conventions of \cite{Buchel:2000cn} and \cite{Pilch:2003jg}. The dilaton $\Phi$ and axion $C_{(0)}$ fields are given by:
\begin{equation}
 e\,^{-\Phi }-i C_{(0)} =\frac{1+\mathcal{B}}{1-\mathcal{B}}\,,
 \qquad
 \mathcal{B}=\,e\,^{2i\phi }\,\frac{\sqrt{cX_1}-\sqrt{X_2}}{\sqrt{cX_1}+\sqrt{X_2}}.
\end{equation}
The 2-form potential that is the linear combination of the RR and the NSNS 2-form potentials, $A_{(2)}=C_{(2)}+i B_{(2)}$, is:
\begin{equation}
 A_{(2)} = e^{i \phi}\left(i a_1 \, d\theta \wedge \sigma_1 + i a_2 \, \sigma_2 \wedge \sigma_3 + a_3\, \sigma_1 \wedge d\phi\right),
\end{equation}
with the real functions\footnote{Notice that we factored out the imaginary $i$, in contrast to \cite{Pilch:2003jg}.}:
\begin{align}
a_1(c,\theta) = & - \frac{\sqrt{c^2-1}}{c}\cos\theta,\nonumber\\
a_2(c,\theta) =  & A  \frac{\sqrt{c^2-1}}{X_1}\sin \theta \cos^2 \theta,\nonumber\\
a_3(c,\theta) =  &  -\frac{\sqrt{c^2-1}}{X_2}\sin \theta \cos^2 \theta.
\end{align}
And the self-dual 5-form field strength $\tilde{F}_{(5)}$ is given by:
\begin{equation}
\tilde{F}_{(5)} = \mathcal{F} + *\mathcal{F},
\qquad
\mathcal{F} = 4 d{x^0} \wedge d{x^1} \wedge d{x^2} \wedge d{x^3} \wedge d\omega(c,\theta),
\end{equation}
where 
\begin{equation}
\omega(c,\theta) = \frac{{A{\kern 1pt} {X_1}}}{{4{{\left( {{c^2} - 1} \right)}^2}}}.
\end{equation}

Using the following definitions for the field strengths:
\begin{align}\label{eq:defs2}
&\tilde F_{(1)} = d C_{(0)},\nonumber\\
&\tilde F_{(3)} = d C_{(2)} + C_{(0)} d B_{(2)},\nonumber\\
&\tilde F_{(5)} = d C_{(4)} + C_{(2)} \wedge d B_{(2)},\nonumber\\
&\tilde F_{(7)} = d C_{(6)} + C_{(4)} \wedge d B_{(2)},\nonumber\\
&\tilde F_{(9)} = d C_{(8)} + C_{(6)} \wedge d B_{(2)},\nonumber\\
\end{align}
and the Hodge duality relation 
\begin{align}\label{eq:dualityconstraint}
\ast \tilde F_{(n+1)} =(-)^{n(n-1)/2} \tilde F_{(9-n)},
\end{align}
we could compute all the RR-forms $C_{(n)}$ and the NSNS-form $B_{(2)}$, up to an exact form. 

We would like to comment on the Hodge star operation, which is defined as:
\begin{equation}
 \ast (dx^{i_1} \wedge \cdots \wedge dx^{i_k} ) = 
 \dfrac{\sqrt{|\det g|}}{(n-k)!} g^{i_1 j_1} \cdots g^{i_k j_k} \epsilon_{j_1 \ldots j_n} (dx^{j_{k+1}} \wedge \cdots \wedge dx^{j_n} ),
\end{equation}
where $g$ is the metric, and the Levi-Civita symbol satisfying $\epsilon_{1 \ldots n}=1$.
Notice the metric to be used here is the metric in the string frame. This is important since the dilaton term is non-trivial in the Pilch-Warner background, unlike in $AdS_5 \times S^5$. 

Although we do not need all the potentials for our problem, we list the explicit solutions below:
\input{sections/forms_explicit}

%% file: sections/forms_explicit.tex
\begin{align*}
e^{\Phi} & =
\frac{c \sin^2\phi \, X_1+\cos^2\phi \,X_2}{\sqrt{c X_1 X_2}}\\
C_{(0)} & = 
-\frac{\sin(2 \phi ) (c X_1-X_2)}{2  (c \sin^2\phi \, X_1+\cos^2\phi \,X_2)}\\
C_{(2)} & = 
- a_2 \sin\phi \,
\sigma_2\wedge \sigma_3
+ a_3 \cos\phi \,
\sigma_1\wedge d\phi
+ a_1 \sin\phi \,
\sigma_1\wedge d\theta\\
B_{(2)} & =
a_2 \cos\phi \,
\sigma_2\wedge \sigma_3
+ a_3 \sin\phi \,
\sigma_1\wedge d\phi
- a_1 \cos\phi \,
\sigma_1\wedge d\theta\\
C_{(4)} & =
4 \, \omega \,
dx_0\wedge dx_1\wedge dx_2\wedge dx_3 
+ \sin\phi \cos\phi \, a_1 a_2 \,
d\theta\wedge \sigma_1\wedge \sigma_2\wedge \sigma_3\\
&
-\left(\cos ^2\phi \, a_2 a_3 +\frac{c \cos^4\theta}{X_2}\right) 
\sigma_1\wedge \sigma_2\wedge \sigma_3 \wedge d\phi \\
C_{(6)} &= - C_{(4)} \wedge B_{(2)}\\
&+\frac{(c A-2) A \sin\theta \cos^2\theta \cos\phi}{2 \left(c^2-1\right)^{5/2}} 
\sigma_1\wedge dc\wedge dx_0\wedge dx_1\wedge dx_2\wedge dx_3\\
&
+\frac{A^2 \cos^3\theta \cos\phi}{2 (c^2-1)^{3/2}}
\sigma_1\wedge dx_0 \wedge dx_1\wedge dx_2\wedge dx_3\wedge d\theta\\
&
-\frac{A^2 \sin\theta \cos^2\theta \sin\phi}{2 (c^2-1)^{3/2}}
\sigma_1\wedge dx_0 \wedge dx_1\wedge dx_2\wedge dx_3\wedge d\phi\\
C_{(8)} &= - C_{(6)} \wedge B_{(2)}\\
&+ \left\{ \frac{A \sin^2\theta \cos^4\theta \cos(2\phi) \left[ ((c^2+3) A-4 c) \cos^2\theta-2 A \right]}{4 (c^2-1)^2 X_2^2} \right.\\
&+ \left.\frac{A ((c^2-1) A + 4 c) \cos^4\theta}{8 c^2 (c^2-1)^2} \right\}
\sigma_1 \wedge \sigma_2\wedge \sigma_3\wedge dc\wedge dx_0\wedge dx_1\wedge dx_2\wedge dx_3\\
& -\frac{A^2 \sin^3(2 \theta ) \cos(2 \phi ) \left(2 c A + A^2 \sin^2\theta +c^2 \cos^2\theta\right)}{16 c (c^2-1) X_2^2}\\
&\sigma_1\wedge \sigma_2\wedge \sigma_3\wedge dx_0\wedge dx_1\wedge dx_2\wedge dx_3\wedge d\theta
\end{align*}

%% file: sections/killing_spinor.tex
Following the appendix of \cite{Chen-Lin:2015xlh}, the Killing spinor solution $\epsilon$ can be written as:
\begin{align}
\epsilon &= V_x^{1/2} \exp{\left(\frac{\phi}{2}\, i \right)} \exp{\left(\frac{\alpha}{2}\Gamma_{56} \right)} \exp{\left(\frac{\beta}{2}\Gamma_{7 10} \mathcal{K} \right)} \eta,
\end{align}
where we defined
\begin{equation}
\eta \equiv \Pi_{+} \mathcal{P}_+ \epsilon_0,
\end{equation}
and the constant spinor $\epsilon_0$ satisfies
\begin{equation}
 \Pi_+ \mathcal{P}_+ \epsilon_0 = \epsilon_0.
\end{equation}

The projectors are
\begin{align}\label{eq:projectors}
\Pi_\pm = 
\frac{1}{2}\left(1 \pm i\Gamma_{1234}\right), 
\quad
\mathcal{P}_{\pm} =
\dfrac{1}{2} \left(1\pm i\Gamma_{6 10}\right),
\end{align}
and they commute with each other: $[\Pi_\pm, \mathcal{P}_{\pm}]=0$.

The exponentials can be written in terms of cosines and sines:
\begin{align}
&\exp{\left(\frac{\alpha}{2}\Gamma_{56} \right)} = \cos\frac{\alpha}{2} + \sin\frac{\alpha}{2}\Gamma_{5 6},\\
&\exp{\left(\frac{\beta}{2}\Gamma_{7 10} \mathcal{K} \right)} = \cos\frac{\beta}{2} + \sin\frac{\beta}{2}\Gamma_{7 10}\mathcal{K},
\end{align}
where $\mathcal{K}$ is the complex conjugation, and the angles are defined as
\begin{align}
\cos\beta = \sqrt{\frac{X_1}{c X_2}}, \quad 
&\quad
\sin\beta = -\sqrt{\frac{c^2 - 1}{c X_2}}\cos\theta,\\
\cos\alpha = \frac{\cos\theta}{\sqrt{X_1}}, 
&\quad
\sin\alpha = \sqrt{\frac{c A}{X_1}}\sin\theta,
\end{align}
where $V_x, X_{1,2}, A$ are functions defined in section \ref{sec:metric}.

The gamma matrices are in the real representations. It is convenient to write the $e^{i \frac{\phi}{2}}$ factor in the real representation too. 
That is achieved by using $\mathcal{P}_- \eta = 0 $, which leads to
\begin{align}
 \exp{\left(\frac{\phi}{2}\, i \right)} \eta 
    = \exp{\left(-\frac{\phi}{2}\Gamma_{6 10} \right)} \eta. 
\end{align}

After some straightforward manipulations, the Killing spinor can be rewritten as:
\begin{equation}\label{eq:KillingSpinor}
\epsilon =  V_x^{1/2} \exp{\left(\frac{\alpha}{2}\Gamma_{56} \right)} \exp{\left(-\frac{\phi}{2}\, \Gamma_{6 10} \right)} \exp{\left(\frac{\beta}{2}\Gamma_{7 10} \mathcal{K} \right)} \Pi_{+} \mathcal{P}_+ \epsilon_0.
\end{equation}

%% file: sections/integrate_action.tex
\section{Integrate on-shell Lagrangian}\label{sec:integrate-action}

The on-shell action \eqref{eq:ActionAtSolution} can be integrated exactly. 
Since it is naturally expanded in powers of $L$, let us define:
\begin{equation}
 S = S_0 + S_2 + S_4.
\end{equation}
The solutions of the respective indefinite integrals, up to constant terms, are:
\begin{align}
 I_0 &= - \int dc \,\left[ \frac{A(c)^2}{4 \left(c^2-1\right)^2}+\frac{c A(c)}{\left(c^2-1\right)^3} \right] \nonumber\\
     &=\frac{A(c)-\left(A(c)^2+2\right) c+3 A(c) c^2}{4 \left(c^2-1\right)^2},\label{eq:I0}\\
    \quad\nonumber\\
I_2 &= - \int dc \, \frac{L^2 A(c) \left(\left(c^2+1\right) A(c)-4 c\right)}{2 \left(c^2-1\right)^2}\nonumber\\
    &=-L^2 \left[\frac{\left(c^3+3 c-4\right) (A(c)-c)^2}{6 \left(c^2-1\right)^2} 
      +\frac{c \left(5-2 c^2\right)}{6 \left(c^2-1\right)} + \, \frac{1}{3} \log^2\left(\frac{c+1}{2}\right)\right.\nonumber\\   
 & \quad \left. +\frac{1}{6} \left(2 A(c)+ c\right)
   + \frac{2}{3}  \, \text{Re} \, \text{Li}_2\left(\frac{c+1}{2}\right)\right],\label{eq:I2}\\
   \quad\nonumber\\
 I_4 &=  \int dc\, \frac{L^4 A(c) \left(3 c^2 A(c)+A(c)-4 c\right)}{4 \left(c^2-1\right)} \nonumber\\
 &= L^4 \left[\frac{\left(9 c^4-7 c^2-2\right) (A(c)-c)}{30 \left(c^2-1\right)}+\frac{\left(9 c^5-10 c^3-15 c+16\right) (A(c)-c)^2}{60 \left(c^2-1\right)^2}\right.\nonumber\\
 & \quad \left. + \frac{3 \, c^3}{20}-\frac{c}{15} -\frac{2}{15} \log ^2\left(\frac{c+1}{2}\right) - \frac{4}{15}\, \text{Re} \, \text{Li}_2\left(\frac{c+1}{2}\right) \right]. \label{eq:I4}
\end{align}

Next, we are going to expand these integrals at different limits. The following expansions for small $\epsilon$ will be useful:
\begin{align}
& \text{Li}_2\left(1 + \epsilon \right) \approx  \zeta(2) + \frac{\pi^2}{4} \epsilon + \ldots,\\
& A(1+\epsilon^2/2)\approx 1+\frac{1}{2} \epsilon^2 \left(2 \log \left(\frac{\epsilon}{2}\right)+1\right) 
+ \frac{1}{8} \epsilon^4 \left(2 \log \left(\frac{\epsilon}{2}\right)-1\right)+ \ldots \label{eq:expandA}
\end{align}

\subsection{UV limit}\label{sec:UVlimit}
Let us evaluate the integrals at the lower bound $c = 1 + \epsilon^2/2$, and expand for small $\epsilon$. The leading order terms are:
\begin{align}
 I_0 &\approx \frac{1}{4 \epsilon ^4} +\frac{1+\log \left(\epsilon/2\right)}{2 \epsilon ^2}-\frac{ \log ^2\left(\epsilon/2\right)}{4}+\frac{\log (\epsilon )}{8} + \frac{3}{64} + O(\epsilon^2),\\
I_2 & \approx -\frac{L^2}{2 \epsilon^2}-\frac{7 L^2}{24}-\frac{\pi ^2 L^2}{9} + O(\epsilon^2),\\
 I_4 & \approx \frac{L^4}{12}-\frac{2 \pi ^2 L^4}{45} + O(\epsilon^2). 
\end{align}

\subsection{IR limit and large L} \label{sec:IRlimit}
Let us evaluate the integrals at $c_\text{max}=\sqrt{1+L^{-2}}$, and expand for large $L$. That means 
\begin{equation}
 c_\text{max} \approx 1 + \frac{1}{2 L^2} - \frac{1}{8 L^4} + O(L^{-6}).
\end{equation}
The leading order terms for the integrals are:
\begin{align}
 I_0 &\approx \frac{L^4}{4}+\frac{1}{2} \left(\frac{5}{4} - \log(2 L)\right) L^2 + O(\log L),\\
I_2 & \approx -\frac{L^4}{2} -\left(\frac{5}{12}+\frac{\pi^2}{9}\right) L^2 + O(\log L),\\
 I_4 & \approx \left(\frac{1}{12}-\frac{2 \pi ^2}{45}\right) L^4+ \frac{1}{120} \left(23 -2 \pi^2 - 44 \log(2 L) \right)L^2+ O(\log L). 
\end{align}
Their sum is:
\begin{equation}\label{eq:IatUV}
 I_0+I_2+I_4 \approx -\left(\frac{1}{6}+\frac{2 \pi ^2}{45}\right) L^4
             +\left(\frac{2}{5}-\frac{23 \pi ^2}{180}-\frac{ 13}{15} \log(2 L)\right)L^2 + O(\log L).
\end{equation}

%% file: main.bbl
%bibliography generated by nb.bst v1.06 (C) 2003-2011 Niklas Beisert
\begin{thebibliography}{10}
\providecommand{\href}[2]{#2}
\providecommand{\arxivref}[2]{\href{http://arxiv.org/abs/#1}{#2}}
\providecommand{\doiref}[2]{\href{http://dx.doi.org/#1}{#2}}
\providecommand{\nbbstauthor}[1]{#1}
\providecommand{\nbbstjournal}[1]{\textsf{#1}}
\providecommand{\nbbsttitle}[1]{\textit{#1}}
\providecommand{\nbbsturl}[1]{\texttt{#1}}
\providecommand{\nbbsteprint}[1]{\texttt{#1}}
\providecommand{\nbbststyle}{\raggedright\small\parskip0pt}
\nbbststyle

\bibitem{Maldacena:1997re}
\nbbstauthor{J.~M.~Maldacena},
\nbbsttitle{``{The Large N limit of superconformal field theories and
  supergravity}''},
\nbbstjournal{\doiref{10.1023/A:1026654312961}{Int.~J.~Theor.~Phys.~38,~1113~(1999)}},
\nbbsteprint{\arxivref{hep-th/9711200}{hep-th/9711200}},
[Adv. Theor. Math. Phys.2,231(1998)].
%%CITATION = HEP-TH/9711200;%%

\bibitem{Karch:2002sh}
\nbbstauthor{A.~Karch and E.~Katz},
\nbbsttitle{``{Adding flavor to AdS / CFT}''},
\nbbstjournal{\doiref{10.1088/1126-6708/2002/06/043}{JHEP~0206,~043~(2002)}},
\nbbsteprint{\arxivref{hep-th/0205236}{hep-th/0205236}}.

\bibitem{CasalderreySolana:2011us}
\nbbstauthor{J.~Casalderrey-Solana, H.~Liu, D.~Mateos, K.~Rajagopal and
  U.~A.~Wiedemann},
\nbbsttitle{``{Gauge/String Duality, Hot QCD and Heavy Ion Collisions}''},
Cambridge University Press (2014).

\bibitem{Erdmenger:2007cm}
\nbbstauthor{J.~Erdmenger, N.~Evans, I.~Kirsch and E.~Threlfall},
\nbbsttitle{``{Mesons in Gauge/Gravity Duals - A Review}''},
\nbbstjournal{\doiref{10.1140/epja/i2007-10540-1}{Eur.~Phys.~J.~A~35,~81~(2008)}},
\nbbsteprint{\arxivref{0711.4467}{arxiv:0711.4467}}.

\bibitem{Pilch:2000ue}
\nbbstauthor{K.~Pilch and N.~P.~Warner},
\nbbsttitle{``{N=2 supersymmetric RG flows and the IIB dilaton}''},
\nbbstjournal{\doiref{10.1016/S0550-3213(00)00656-8}{Nucl.~Phys.~B594,~209~(2001)}},
\nbbsteprint{\arxivref{hep-th/0004063}{hep-th/0004063}}.
%%CITATION = HEP-TH/0004063;%%

\bibitem{Pilch:2003jg}
\nbbstauthor{K.~Pilch and N.~P.~Warner},
\nbbsttitle{``{Generalizing the N=2 supersymmetric RG flow solution of IIB
  supergravity}''},
\nbbstjournal{\doiref{10.1016/j.nuclphysb.2003.09.052}{Nucl.~Phys.~B675,~99~(2003)}},
\nbbsteprint{\arxivref{hep-th/0306098}{hep-th/0306098}}.
%%CITATION = HEP-TH/0306098;%%

\bibitem{Buchel:2013id}
\nbbstauthor{A.~Buchel, J.~G.~Russo and K.~Zarembo},
\nbbsttitle{``{Rigorous Test of Non-conformal Holography: Wilson Loops in N=2*
  Theory}''},
\nbbstjournal{\doiref{10.1007/JHEP03(2013)062}{JHEP~1303,~062~(2013)}},
\nbbsteprint{\arxivref{1301.1597}{arxiv:1301.1597}}.
%%CITATION = ARXIV:1301.1597;%%

\bibitem{Chen-Lin:2015xlh}
\nbbstauthor{X.~Chen-Lin, A.~Dekel and K.~Zarembo},
\nbbsttitle{``{Holographic Wilson loops in symmetric representations in N = 2*
  super-Yang-Mills theory}''},
\nbbstjournal{\doiref{10.1007/JHEP02(2016)109}{JHEP~1602,~109~(2016)}},
\nbbsteprint{\arxivref{1512.06420}{arxiv:1512.06420}}.
%%CITATION = ARXIV:1512.06420;%%

\bibitem{Chen-Lin:2017pay}
\nbbstauthor{X.~Chen-Lin, D.~Medina-Rincon and K.~Zarembo},
\nbbsttitle{``{Quantum String Test of Nonconformal Holography}''},
\nbbstjournal{\doiref{10.1007/JHEP04(2017)095}{JHEP~1704,~095~(2017)}},
\nbbsteprint{\arxivref{1702.07954}{arxiv:1702.07954}}.
%%CITATION = ARXIV:1702.07954;%%

\bibitem{Russo:2019lgq}
\nbbstauthor{J.~G.~Russo, E.~Widén and K.~Zarembo},
\nbbsttitle{``{N = 2* phase transitions and holography}''},
\nbbstjournal{\doiref{10.1007/JHEP02(2019)196}{JHEP~1902,~196~(2019)}},
\nbbsteprint{\arxivref{1901.02835}{arxiv:1901.02835}}.

\bibitem{Albash:2011nw}
\nbbstauthor{T.~Albash and C.~V.~Johnson},
\nbbsttitle{``{Dynamics of Fundamental Matter in N=2* Yang-Mills Theory}''},
\nbbstjournal{\doiref{10.1007/JHEP04(2011)012}{JHEP~1104,~012~(2011)}},
\nbbsteprint{\arxivref{1102.0554}{arxiv:1102.0554}}.

\bibitem{Skenderis:2002vf}
\nbbstauthor{K.~Skenderis and M.~Taylor},
\nbbsttitle{``{Branes in AdS and p p wave space-times}''},
\nbbstjournal{\doiref{10.1088/1126-6708/2002/06/025}{JHEP~0206,~025~(2002)}},
\nbbsteprint{\arxivref{hep-th/0204054}{hep-th/0204054}}.
%%CITATION = HEP-TH/0204054;%%

\bibitem{Kruczenski:2003be}
\nbbstauthor{M.~Kruczenski, D.~Mateos, R.~C.~Myers and D.~J.~Winters},
\nbbsttitle{``{Meson spectroscopy in AdS / CFT with flavor}''},
\nbbstjournal{\doiref{10.1088/1126-6708/2003/07/049}{JHEP~0307,~049~(2003)}},
\nbbsteprint{\arxivref{hep-th/0304032}{hep-th/0304032}}.

\bibitem{Ammon:2015wua}
\nbbstauthor{M.~Ammon and J.~Erdmenger},
\nbbsttitle{``{Gauge/gravity duality}''},
Cambridge Univ. Pr. (2015),
Cambridge, UK.

\bibitem{Karch:2005ms}
\nbbstauthor{A.~Karch, A.~O'Bannon and K.~Skenderis},
\nbbsttitle{``{Holographic renormalization of probe D-branes in AdS/CFT}''},
\nbbstjournal{\doiref{10.1088/1126-6708/2006/04/015}{JHEP~0604,~015~(2006)}},
\nbbsteprint{\arxivref{hep-th/0512125}{hep-th/0512125}}.

\bibitem{Buchel:2000cn}
\nbbstauthor{A.~Buchel, A.~W.~Peet and J.~Polchinski},
\nbbsttitle{``{Gauge dual and noncommutative extension of an N=2 supergravity
  solution}''},
\nbbstjournal{\doiref{10.1103/PhysRevD.63.044009}{Phys.~Rev.~D63,~044009~(2001)}},
\nbbsteprint{\arxivref{hep-th/0008076}{hep-th/0008076}}.
%%CITATION = HEP-TH/0008076;%%

\bibitem{Evans:2000ct}
\nbbstauthor{N.~J.~Evans, C.~V.~Johnson and M.~Petrini},
\nbbsttitle{``{The Enhancon and N=2 gauge theory: Gravity RG flows}''},
\nbbstjournal{\doiref{10.1088/1126-6708/2000/10/022}{JHEP~0010,~022~(2000)}},
\nbbsteprint{\arxivref{hep-th/0008081}{hep-th/0008081}}.

\bibitem{Evans:2005ti}
\nbbstauthor{N.~Evans, J.~P.~Shock and T.~Waterson},
\nbbsttitle{``{D7 brane embeddings and chiral symmetry breaking}''},
\nbbstjournal{\doiref{10.1088/1126-6708/2005/03/005}{JHEP~0503,~005~(2005)}},
\nbbsteprint{\arxivref{hep-th/0502091}{hep-th/0502091}}.

\bibitem{Bobev:2013cja}
\nbbstauthor{N.~Bobev, H.~Elvang, D.~Z.~Freedman and S.~S.~Pufu},
\nbbsttitle{``{Holography for N = 2* on S4}''},
\nbbstjournal{\doiref{10.1007/JHEP07(2014)001}{JHEP~1407,~001~(2014)}},
\nbbsteprint{\arxivref{1311.1508}{arxiv:1311.1508}}.

\bibitem{Bergner:2014saa}
\nbbstauthor{G.~Bergner, P.~Giudice, G.~Münster, S.~Piemonte and
  D.~Sandbrink},
\nbbsttitle{``{Phase structure of the N=1 supersymmetric Yang-Mills theory at
  finite temperature}''},
\nbbstjournal{\doiref{10.1007/JHEP11(2014)049}{JHEP~1411,~049~(2014)}},
\nbbsteprint{\arxivref{1405.3180}{arxiv:1405.3180}}.

\bibitem{Karch:2015vra}
\nbbstauthor{A.~Karch, B.~Robinson and C.~F.~Uhlemann},
\nbbsttitle{``{Supersymmetric D3/D7 for holographic flavors on curved
  space}''},
\nbbstjournal{\doiref{10.1007/JHEP11(2015)112}{JHEP~1511,~112~(2015)}},
\nbbsteprint{\arxivref{1508.06996}{arxiv:1508.06996}}.

\bibitem{Albash:2011dq}
\nbbstauthor{T.~Albash and C.~V.~Johnson},
\nbbsttitle{``{Thermal Dynamics of Quarks and Mesons in N=2* Yang-Mills
  Theory}''},
\nbbstjournal{\doiref{10.1007/JHEP07(2011)063}{JHEP~1107,~063~(2011)}},
\nbbsteprint{\arxivref{1105.3202}{arxiv:1105.3202}}.

\bibitem{Buchel:2003ah}
\nbbstauthor{A.~Buchel and J.~T.~Liu},
\nbbsttitle{``{Thermodynamics of the N=2* flow}''},
\nbbstjournal{\doiref{10.1088/1126-6708/2003/11/031}{JHEP~0311,~031~(2003)}},
\nbbsteprint{\arxivref{hep-th/0305064}{hep-th/0305064}}.

\end{thebibliography}
